\documentstyle[twocolumn,psbox]{jpsj}

\def\runtitle{
Origin of Superconductivity in 2D-Organic Compounds and High-$T_{\rm c}$ Cuprates
}
\def\runauthor{Hisashi {\sc Kondo} and T\^{o}ru {\sc Moriya}}

\title
{
Origin of Superconductivity \\ 
in 2D-Organic Compounds and High-$T_{\rm c}$ Cuprates
}

\author
{ 
Hisashi {\sc Kondo} and T\^{o}ru {\sc Moriya}
}

\inst
{
Department of Physics, 
Faculty of Science and Technology, Science University of Tokyo, Noda 278-8510
}

\recdate
{\today}

\abst
{
     A phase diagram is drawn in a parameter space of the nearly half-filled 
single band two-dimensional Hubbard model with $U/t$, $t^{\prime}/t$ and $n$ 
as the parameters, $U$, $t$, $t^{\prime}$ and $n$ being the on-site interaction, 
the nearest and second nearest neighbor transfer matrices and the number of 
electrons per site, respectively. 
The parameter space is chosen so as to include the models 
for the high-$T_{\rm c}$ cuprates as well as 2D-organic superconductors. 
The superconducting and antiferromagnetic instability surfaces are drawn from 
the results of calculations by using the spin fluctuation theory within the 
fluctuation exchange approximation. The metal-insulator transition surface is 
also indicated in the diagram. From this phase diagram we discuss the 
possibility of the spin fluctuation mechanism being the common origin of the 
superconductivity in the high-$T_{\rm c}$ cuprates and 2D-organic compounds. 
}

\kword
{
superconductivity, high-$T_{\rm c}$ cuprates, organic compounds, 
spin fluctuations, Hubbard model, pseudo-spin gap
}

\begin{document}
\sloppy
\maketitle


     The origin of superconductivity in strongly correlated electron systems 
has been the central subject of current interest in condensed matter physics. 
It is widely believed that the spin fluctuation mechanism is responsible for 
the superconductivity in the heavy electron systems. On the other hand, the 
mechanism for the high-$T_{\rm c}$ cuprates is still under controversy. 
Although the magnetic origin seems to be widely accepted, a long-standing 
controversy still persists between the spin fluctuation (SF) 
mechanism~\cite{rf:1,rf:2,rf:3} 
and the resonating valence bond (RVB) mechanism based on the $t$-$J$ 
model.~\cite{rf:4,rf:5,rf:6,rf:32}
 
     Recently, the superconductivity in two-dimensional (2D) organic conductors: 
$\kappa$-$({\rm ET})_2 X$  [${\rm ET} = {\rm BEDT}$-${\rm TTF}$, 
$X = {\rm Cu \{ N (CN)_2 \} } X^{\prime}, X^{\prime} = {\rm Cl}, {\rm Br}, ...$] 
has been successfully investigated in terms of the spin fluctuation 
theory.~\cite{rf:7,rf:8,rf:9,rf:10,rf:11,rf:12}
The conducting electrons in these systems are believed to be well described in 
terms of a half-filled 2D-single band Hubbard model consisting of antibonding 
dimer orbitals. Thus the spin fluctuation mechanism seems to be the only 
available mechanism which explains the anisotropic superconductivity in these 
substances;~\cite{rf:13,rf:14,rf:15} 
the $t$-$J$ model as applied to a half-filled band leads 
to nothing but an insulating state. 

     As for the cuprates, investigations in this decade have indicated that the 
essential physics is well described in terms of the 2D-single 
band Hubbard model with the transfer matrices up to the second or third 
nearest neighbors in the ${\rm Cu O}_2$ plane. The more realistic $d$-$p$ 
model consisting of ${\rm Cu}$-$d(x^2-y^2)$ and ${\rm O}$-$p \sigma$ 
orbitals is considered to lead to essentially the same results as the Hubbard 
model, or to be reduced, after proper renormalizations of high frequency 
components, to an effective single band Hubbard model.~\cite{rf:16} 
It should be noted that the 
$t$-$J$ model is derived from these models as the lowest order expansion 
from the strong coupling limit. Since the above-mentioned organic compounds 
are also well described in terms of the Hubbard model, we expect that these 
two classes of substances can be put in the same phase diagram by a proper 
choice of the parameter space for the 2D-Hubbard model. 
\begin{figure}[t]
  \begin{center}
    \psbox[size=1.#1]{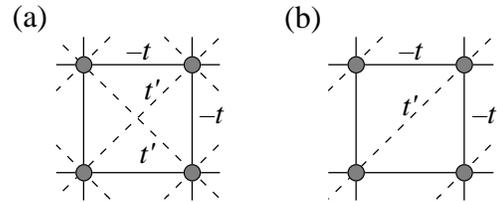}
  \end{center}
  \caption{
     Transfer matrices in the 2D-Hubbard model 
     (a) for cuprates and (b) for $\kappa$-$({\rm ET})_2 X$, 
     2D-organic superconductors.
  }
\end{figure}

     Let us consider a square lattice for both cuprates and organic compounds. 
The nearest neighbor transfer matrix is given by $-t$ $( t > 0)$. 
The second nearest neighbor transfer matrix $t^{\prime}$ for cuprates as 
shown in Fig. 1(a) 
takes various values; the commonly used values are 
$(t^{\prime}/t) \approx 0.2$ for ${\rm La}_{1-x} {\rm Sr}_x {\rm CuO}_4$ 
and $\approx 0.5$ for ${\rm YBa}_2 {\rm Cu}_3 {\rm O}_{6+x}$. For brevity 
we neglect here the transfer matrices beyond the second neighbors. For 
$\kappa$-$({\rm ET})_2 X$ the simplest and good model is to take only two out 
of the four second neighbor transfers $t^{\prime}$, as shown in Fig. 1(b). 
The estimated value for this transfer matrix is $(t^{\prime}/t) \approx -0.8$. 
The occupation number $n$ of electrons per site is $1$, or the band is 
half-filled, for $\kappa$-$({\rm ET})_2 X$ and $< 1$ for hole-doped cuprates, 
best studied high-$T_{\rm c}$ substances. So it seems illuminating to draw 
a phase diagram in a three dimensional parameter space with $n$ and 
$t^{\prime}/t$ for the horizontal axes and $U/t$ for the vertical axis, 
taking the transfer matrices in Figs. 1(a) and 1(b) for $t^{\prime}/t > 0$ and 
$t^{\prime}/t < 0$, respectively. For this purpose we must calculate the 
phase diagrams in various two-dimensional planes, $U/t$ against $n$ for various 
fixed values of $t^{\prime}/t$, and/or $U/t$ against 
$t^{\prime}/t$ for various fixed values of $n$. 

     For this purpose, the information available on the ground state of the 
Hubbard model is extremely limited. Let us discuss the phase diagram 
calculated on the basis of the self-consistent theory of coupled modes of 
spin fluctuations as applied to magnetism and spin fluctuation-induced 
superconductivity. In practice, we use the simplest practicable one, i.e., 
the so-called fluctuation exchange (FLEX) 
approximation.~\cite{rf:17,rf:18,rf:19,rf:20,rf:21,rf:22,rf:23} 
This approximation with no vertex correction has been successfully 
applied to the cuprates and the 2D-organic superconductors and recent 
studies on the vertex corrections reveal that they do not seriously affect 
the FLEX results.~\cite{rf:20,rf:24,rf:25} 
Since the numerical work of this type of theory 
is usually carried out at finite temperatures, we need to extrapolate 
the results to zero temperature. 

    In our previous study of $\kappa$-$({\rm ET})_2 X$ we showed such a 
phase diagram for $U/t$ against the negative axis of $t^{\prime}/t$ for $n = 1$; 
the phase boundary between the paramagnetic metallic (PM) and 
superconducting (SC) phases was obtained by extrapolating $T_{\rm c}$ 
against $U/t$ 
curves to $T_{\rm c} = 0$ for various values of $t^{\prime}/t$ with the use of 
the Pad\'{e} approximants.~\cite{rf:8} 
A similar extrapolation of the $T_{\rm c}$ against $n$ 
curve to $T_{\rm c} = 0$ for various values of $U/t$ leads to the PM-SC phase 
boundary in the $U/t$ against $n$ plane for a given value of $t^{\prime}/t$. 
The antiferromagnetic (AF) instability line or the PM-AFM (AF metal) 
phase boundary was obtained by extrapolating the inversed staggered 
susceptibility $1/\chi_{\mibs Q}$ to $T = 0$ and picking up the parameter 
values giving $1/\chi_{\mibs Q} (T=0) =0$. 

     In order to construct a three-dimensional phase diagram as discussed 
above we must extend this type of calculation to include the models with 
various values of $t^{\prime}/t$ and $n$. The phase diagram for 
$t^{\prime}/t > 0$ is smoothly connected with that for $t^{\prime}/t < 0$, 
with only half the second neighbor transfers, through the $t^{\prime}/t = 0$ 
plane. The former includes the cuprates and the latter includes 
$\kappa$-$({\rm ET})_2 X$.  We have calculated several phase diagrams as 
shown in Fig. 2, 
enabling the construction of a three-dimensional phase diagram, as 
sketched in Fig. 3. 
In Figs. 2 and 3 we adopt $10 > U/t >0$ for the range of parameter 
values which is expected to cover both the cuprates and $\kappa$-$({\rm ET})_2 X$.
\begin{figure}[t]
  \begin{center}
    \psbox[size=0.975#1]{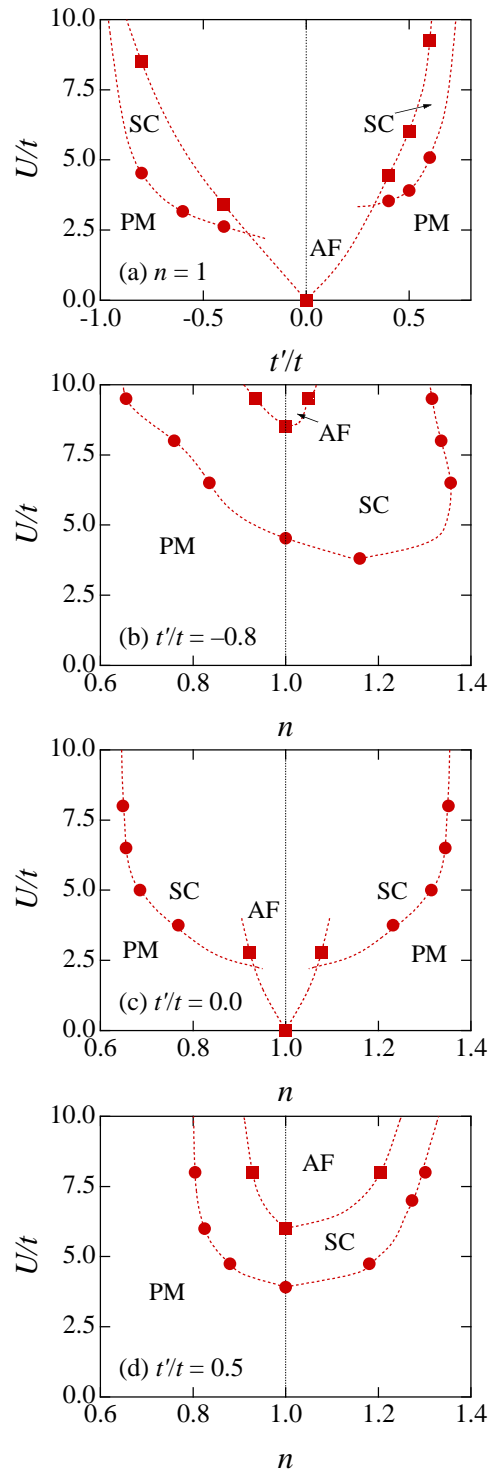}
  \end{center}
  \caption{
     Superconducting and antiferromagnetic instability lines in
     (a) $U/t$ against $t^{\prime}/t$ plane for $n = 1$ and in (b), (c) and 
     (d) $U/t$ against $n$ plane for $t^\prime/t = -0.8$, $0$ and $0.5$.
  }
\end{figure}
\begin{figure}[t]
  \begin{center}
    \psbox[size=.975#1]{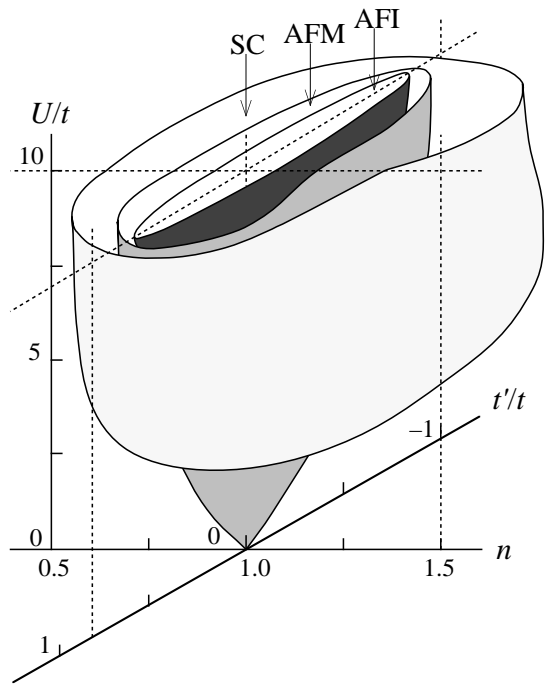}
  \end{center}
  \caption{
     Phase diagram in a three-dimensional 
     ($U/t$)-($t^{\prime}/t$)-$n$ space. Antiferromagnetic metal to insulator 
     transition surface which should depend on the impurity potential is 
     sketched here to give a general idea of its location.
     Phase boundary surfaces are cut at the plane of $U/t=10$.
  }
\end{figure}

     Although this sketch is by no means accurate, the essential features of 
the phase diagram are considered to be reproduced correctly. 
The PM-SC phase boundary surface is a real one and should actually be 
observed. On the other hand, the antiferromagnetic instability surface, 
separating the PM and AFM phases is realistic when it is outside the PM-SC 
boundary surface, as is seen in the lower part of Fig. 3, but unrealistic 
when it is inside it, as is seen in the upper part of Fig. 3.  
In the latter case the SC phase should extend beyond the PM-AFM boundary, 
since the SC phase is more stable than the AFM phase around this boundary. 
Thus we should find an SC-AF boundary surface, probably of first order 
transition. This surface, which is still to be calculated by comparing the 
energies of the two phases, is not shown in Fig. 3.  
The possibility of SC-AF coexistence is left for future investigations.
The antiferromagnetic 
insulator (AFI) phase exist in the $n = 1$ plane above an AFM-AFI boundary 
line located usually above the PM-AFM boundary line. 
In reality, the AFI phase extends 
into the $n \neq 1$ region owing to the Anderson localization due to 
the unavoidable impurity potential. Considering this effect an AFM-AFI 
boundary surface is sketched arbitrarily in Fig. 3 as an aid to  
understanding the overall picture. 

     From this figure we find that there is a substantial volume of 
superconducting phase outside the surface of antiferromagnetic instability 
in a range of intermediate size of $U/t$ or $U/W \sim 1$, $W \approx 8t$ 
being the bandwidth. As was mentioned previously, the superconducting phase 
for $n = 1$ (half-filled) can be understood only in terms of the spin 
fluctuation mechanism and this phase on both sides of 
the $t^{\prime}/t = 0$ plane
smoothly continues to the electron-doped ($n > 1$) and hole-doped ($n < 1$) 
phases. The $\kappa$-$({\rm ET})_2$ compounds as well as high-$T_{\rm c}$ 
cuprates find their positions within the one connected region of this phase 
diagram. This seems to indicate very strongly that these superconductors are 
consistently explained in terms of the spin fluctuation mechanism as was 
shown within the FLEX approximation.
     
     We have seen that the SC phase extends into the AFM region. 
It is interesting to note that the pseudo-gap phenomena, both in the 
spin excitation spectrum and in the one electron spectral density, are 
observed in this regime both in cuprates and in 
$\kappa$-$({\rm ET})_2 X$.~\cite{rf:13,rf:14,rf:15,rf:26} 
Since these phenomena do not seem to be 
described by the FLEX approximation we must improve the theory in this 
regime. Leaving discussions on the possible origins of the pseudo-gap 
phenomena for later part of this article, we may conclude for the moment 
that the SC phase outside this pseudo-gap regime is due to the 
antiferromagnetic spin fluctuations and that this SC phase smoothly extends into 
the pseudo-gap regime.
     
     Let us next consider to what extent 
the $t$-$J$ model can cover this phase diagram. 
As mentioned previously, this model can never deal with the 
superconductivity for $n = 1$ as in $\kappa$-$({\rm ET})_2$ compounds. 
It is also hard to deal with the SC phase outside the AFM 
regime of relatively small $U/t$, 
at least below the critical value $(U/t)_{\rm c}$ giving rise to an AFI phase. 
This part of the SC phase is well 
described by the spin fluctuation theory. What is clear is that the $t$-$J$ 
model has a possibility of covering only a limited range of relatively 
large $U/t$ and small $\delta n=|n-1|$ in the phase diagram. 
     
     The $t$-$J$ model assumes an infinite value of $U/t$, strictly 
forbidding double occupancy of a site. Although a larger value of $J$ 
corresponds to a smaller value of $U/t$, the complete exclusion of double 
occupancy makes it hard to judge if one can really approach a realistic 
intermediate range of values for $U/t$. As a matter of fact a typical 
value of $J/t$ for the $t$-$J$ model for the cuprates seems to be 
$\sim 0.5$, corresponding to $U/t = 8$ in the Hubbard model of the intermediate 
coupling regime. According to the variational Monte Carlo calculations for 
the superconducting condensation energy of these models, the Hubbard model 
with $U/t \sim 8$, $t^{\prime}=0$, $n=0.84$ 
gives a value consistent with experiments, while the corresponding $t$-$J$ 
model with $J/t \sim 0.5$ gives a value more than twenty times larger than 
the observed value.~\cite{rf:27} 
This result seems to pose a serious problem for the $t$-$J$ model which is an 
approximate form of the Hubbard model for large $U/t$ and small $\delta n=|n-1|$.
It seems that the part of the parameter space covered 
by the $t$-$J$ model does not extend sufficiently far to overlap the part 
studied here 
which is expected to contain the high-$T_{\rm c}$ cuprates 
as well as 2D-organic superconductors. 
     
     Now let us discuss the pseudo-spin gap phenomena occurring in the 
transition region between the antiferromagnetic insulator phase and the 
superconducting phase.~\cite{rf:26} 
Although there are several scenarios presented 
to explain these phenomena, none of them seems to be completely 
satisfactory. Recent 
observations of various pseudo-gap behaviors in $\kappa$-$({\rm ET})_2 X$, 
quite similar to those in underdoped 
cuprates~\cite{rf:13,rf:14,rf:15,rf:28,rf:29} 
indicate strongly that the origin of the phenomena is common to 
both of these systems. 
     
     We stress here that the possible mechanism to explain the pseudo-gaps 
in $\kappa$-$({\rm ET})_2 X$ is naturally expected to cover those in cuprates, 
too. On the other hand, the $t$-$J$ model mechanism or the spinon 
pairing mechanism 
for the pseudo-gaps cannot cover $\kappa$-$({\rm ET})_2 X$. 
Thus it seems wise not to pursue the $t$-$J$ model scenario for the pseudo-gaps 
before the range of validity of the model itself is clarified. 
In doing so we can avoid looking for two different mechanisms 
for the same phenomena. 
     
     The other proposed mechanisms include the spin density wave (SDW) 
fluctuations, the coexistence of charge and spin density wave (CDW and SDW) 
fluctuations and the influence of superconducting fluctuations or formation 
of preformed pairs above $T_{\rm c}$. 
We note that these mechanisms for the pseudo-gaps, 
the validly of which is still to be examined, necessarily assume the 
spin fluctuation mechanism as the origin of the d-wave superconductivity. 

     As a consequence of the 2D-antiferromagnetic spin fluctuation 
theory we expect to have 2D-SDW fluctuations of very low frequencies beyond 
the AF instability point. This seems to give a natural explanation for the 
pseudo-gap in one-electron spectral density. A difficulty with this mechanism 
is that it is incapable of explaining magnetic excitation gaps or the temperature 
dependence of the nuclear spin-lattice relaxation rate. Insofar as we have 
an AFM or SDW ground state (when the SC phase is suppressed) this difficulty 
seems unavoidable. One solution may be to consider 
additional charge density wave (CDW) fluctuations which provide a magnetic 
excitation gap. Although this is an appealing possibility for both cuprates 
and organic compounds, it is not clear if the CDW fluctuations can be so 
universal as the pseudo-gap phenomena in these systems.      
     
     We may think of another possible way out of this difficulty. 
Unfortunately the AFM phase 
in a single band 2D-Hubbard model has not been well enough studied beyond 
the mean field and FLEX approximations. Even the stability of the metallic 
AF ground state in this case has not been proved either theoretically or 
experimentally. Possibly the necessarily small 
AF moment ($<1/2$) is not sustained in this 2D-metallic phase with large 
quantum fluctuations, leading to a non-magnetic ground state with a pseudo 
SDW and magnetic excitation gaps. 
The pseudo-SDW gap formation would naturally suppress the superconducting gap, 
leading to an explanation for the doping concentration dependence of 
$T_{\rm c}$ in the underdoped regime. 
The key issue here is to explain the 
magnetic excitation gap, possibly by local singlet correlations or magnetic 
anisotropy. Although this is simply conjecture to be substantiated later, 
the influence of this possible mechanism on the spin fluctuation-induced 
superconductivity will not be large since high frequency components 
of the spin fluctuations mainly contribute to the 
superconductivity.~\cite{rf:30,rf:31}
     
     Lastly, as a consequence of the above arguments we suggest that it 
seems unlikely that the $t$-$J$ model explanation 
for the high-$T_{\rm c}$ superconductivity in cuprates and the spinon pairing 
mechanism for the spin pseudo-gap phenomena are realistic. 
Further careful examination of the range of validity of 
the $t$-$J$ model is desired.
     
     We may now conclude with high probability that the superconductivity of the 
organic 2D-compounds as well as that of the high-$T_{\rm c}$ cuprates is 
consistently explained in terms of the spin fluctuation 
mechanism. Although we still have one serious problem of explaining the 
pseudo-gap phenomena observed in the underdoped cuprates and in the 
organic 2D-compounds, the main physics of the spin fluctuation-induced 
superconductivity is not expected to be much influenced by the 
underlying mechanism of these phenomena. 

     We would like to thank K. Ueda, S. Nakamura and T. Takimoto for helpful 
discussions and Y. Nakazawa and K. Nomura for useful communications.

\end{document}